\documentclass[reprint, amsmath, amssymb, aps, superscriptaddress, nofootinbib, nobibnotes,onecolumn]{revtex4}

\pdfoutput=1
\usepackage{graphicx}
\graphicspath{{figures/}}

\usepackage[utf8]{inputenc}
\usepackage{amsmath}
\usepackage{amsfonts}
\usepackage{amssymb}
\usepackage{multirow}
\usepackage[colorlinks=true, linkcolor=red, citecolor=blue]{hyperref}
\usepackage[capitalise]{cleveref}
\crefname{figure}{Fig.}{Figs.}
\Crefname{figure}{Fig.}{Figs.}

\def\({\left(}
\def\){\right)}
\def\[{\left[}
\def\]{\right]}
\def\be{\begin{eqnarray}}
\def\ee{\end{eqnarray}}

\usepackage{acro}
\DeclareAcronym{BH}{
	short = BH ,
	long  = black hole
}
\DeclareAcronym{PBH}{
	short = PBH ,
	long  = primordial black hole
}
\DeclareAcronym{BBH}{
	short = BBH ,
	long  = binary black hole
}
\DeclareAcronym{GW}{
	short = GW ,
	long  = gravitational wave
}
\DeclareAcronym{SGWB}{
	short = SGWB ,
	long  = stochastic gravitational-wave background
}
\DeclareAcronym{RD}{
	short = RD ,
	long  = radiation dominated
} 
\DeclareAcronym{CMB}{
	short = CMB ,
	long  = cosmic microwave background
} 
\DeclareAcronym{CDM}{
	short = CDM ,
	long  = cold dark matter
}
\DeclareAcronym{SNR}{
	short = SNR ,
	long  = signal-to-noise ratio
}
\DeclareAcronym{aLIGO}{
	short = aLIGO ,
	long  = Advanced LIGO
}

\DeclareAcronym{SKA}{
	short = SKA ,
	long  = Square Kilometre Array
}
\DeclareAcronym{LISA}{
	short = LISA ,
	long  = Laser Interferometer Space Antenna
}
\DeclareAcronym{eLISA}{
	short = eLISA ,
	long  = evolved LISA
}
\DeclareAcronym{DECIGO}{
	short = DECIGO ,
	long  = DECi-hertz Interferometer Gravitational wave Observatory
}
\DeclareAcronym{BBO}{
	short = BBO ,
	long  = Big Bang Observer
}
\DeclareAcronym{ET}{
	short = ET ,
	long  = Einstein Telescope
}
\DeclareAcronym{WIMP}{
	short = WIMP ,
	long  = weakly interacting massive particle
}

\begin{document}

\title{Prospective constraints on the primordial black hole abundance \\ from the stochastic gravitational-wave backgrounds  \\ produced by coalescing events and curvature perturbations}

\author{Sai Wang}
\email{wangsai@post.kek.jp}
\affiliation{Institute of Particle and Nuclear Studies, KEK, 1-1 Oho, Tsukuba 305-0801, Japan}
\author{Takahiro Terada}
\email{teradat@post.kek.jp}
\affiliation{Institute of Particle and Nuclear Studies, KEK, 1-1 Oho, Tsukuba 305-0801, Japan}
\author{Kazunori Kohri}
\email{kohri@post.kek.jp}
\affiliation{Institute of Particle and Nuclear Studies, KEK, 1-1 Oho, Tsukuba 305-0801, Japan}
\affiliation{The Graduate University for Advanced Studies (SOKENDAI), 1-1 Oho, Tsukuba 305-0801, Japan}

%%%%%%%%%%%%%%%%%%%%%%%%%%%%%%%%%%%%%%%%%%%%%%%%%%%%%%%%%%%%%%%%%%%%%%
\begin{abstract}
\noindent 
For a variety of on-going and planned gravitational-wave (GW)
experiments, we study expected constraints on the fraction
($f_{\rm PBH}$) of primordial black holes (PBHs) in
dark matter by evaluating the energy-density spectra of two kinds of
stochastic GW backgrounds.  The first one is produced from an
incoherent superposition of GWs emitted from coalescences of all the
binary PBHs. The second one is induced through non-linear mode
couplings of large primordial curvature perturbations
inevitably associated with the generation of PBHs in the early
Universe.  In this paper, we focus on the PBHs with their masses of
$10^{-8}M_{\odot}\leq M_{\mathrm{PBH}} < 1M_{\odot}$, since they are
not expected to be of a stellar origin. In almost all ranges of the masses, we show that the experiments are sensitive to constrain the
fraction for $10^{-5} \lesssim f_{\rm PBH} \lesssim 1$ by considering the GWs from coalescing events and $10^{-13} \lesssim f_{\rm PBH} \lesssim 1$ by considering the GWs from curvature perturbations.  Exceptionally, only in a narrow range of masses for
$M_{\mathrm{PBH}} \simeq 10^{-7} M_{\odot}$,
the fraction cannot be constrained for $f_{\rm PBH} \lesssim 10^{-13}$ by those two GW backgrounds.
\end{abstract}
%%%%%%%%%%%%%%%%%%%%%%%%%%%%%%%%%%%%%%%%%%%%%%%%%%%%%%%%%%%%%%%%%%%%%%

\maketitle
%%%%%%%%%%%%%%%%%%%%%%%%%%%%%%%%%%%%%%%%
%%%%%%%%%%%%%%%%%%%%%%%%%%%%%%%%%%%%%%%
\acresetall

%%%%%%%%%%%%%%%%%%%%%%%%%%%%%%%%%%%%%%%%%%%%%%%%%%%%%%%%%%%%%%%%%%%%%%
\section{Introduction}
%%%%%%%%%%%%%%%%%%%%%%%%%%%%%%%%%%%%%%%%%%%%%%%%%%%%%%%%%%%%%%%%%%%%%%
\noindent
The first detection of \acp{GW} from a \ac{BBH} merger by the first \ac{aLIGO} observing run \cite{Abbott:2016blz} has revived extensive interests in \acp{PBH} \cite{Hawking:1971ei,Carr:1974nx}, which are produced 
directly from the gravitational collapses of the enhanced inhomogeneities in the primordial Universe. Until now, the origin of these \acp{BH} and the formation mechanism of \acp{BBH} are still under debate. Besides an astrophysical origin \cite{TheLIGOScientific:2016htt,Belczynski:2016obo,Miller:2016krr}, the possibility that these \acp{BH} are of a primordial origin is also considered \cite{Bird:2016dcv,Sasaki:2016jop,Clesse:2016vqa,Eroshenko:2016hmn,Carr:2016drx,Kashlinsky:2016sdv,Bartolo:2016ami,Cholis:2016xvo,Harada:2016mhb,Georg:2017mqk,Nakamura:2016hna,Raccanelli:2016cud,Carr:2017jsz,Sasaki:2018dmp}. Recently, it has been proposed that the \acp{PBH} are capable of accounting for the event rate of \ac{BBH} mergers observed by \ac{aLIGO} \cite{Bird:2016dcv,Sasaki:2016jop}, although the formation mechanisms of \ac{PBH} binaries bring about uncertainties of a couple of orders of magnitude (see e.g. Ref.~\cite{Sasaki:2018dmp} and references therein).
The \acp{PBH} can be one of the most promising candidates for the \ac{CDM} \cite{Carr:2016drx}. Currently, the nature of
\ac{CDM} is still uncertain \cite{Bertone:2016nfn}. There is not
definitive evidence for the \acp{WIMP} which is a prime candidate for
\ac{CDM}
\cite{Tan:2016zwf,Akerib:2015rjg,ATLAS:2016hao,Accardo:2014lma,FermiLAT:2011ab}. Conventionally,
one defines the abundance of \acp{PBH} in \ac{CDM} as a dimensionless
fraction of the form
$f_{\mathrm{PBH}}=\Omega_{\mathrm{PBH}}/\Omega_{\mathrm{CDM}}$, where
$\Omega_{\mathrm{PBH}}$ and $\Omega_{\mathrm{CDM}}$ denote the present
energy-density fractions of \acp{PBH} and \ac{CDM}, respectively. This
quantity has been constrained in a variety of mass ranges by a variety
of observations (see e.g. Refs.~\cite{Carr:2009jm,Sasaki:2018dmp} and
references therein), for example, the microlensing events caused by
massive astrophysical compact halo objects
\cite{Novati:2013fxa,Mediavilla:2009um,Green:2016xgy,Axelrod:2016nkp},
the gas accretion effect of \acp{PBH} on the \ac{CMB}
\cite{Chen:2016pud,Ali-Haimoud:2016mbv,Poulin:2017bwe}, the null
detection of a third-order Shapiro time delay using a pulsar timing
array \cite{Schutz:2016khr}, and the claimed event rate of \ac{BBH}
mergers from \ac{aLIGO}
\cite{Bird:2016dcv,Sasaki:2016jop,Abbott:2018oah}, etc.

The \acp{PBH} can be also a useful probe to the primordial curvature
perturbations \cite{Mukhanov:1990me}, since the former are formed via
directly gravitational collapses of the latter
\cite{Hawking:1971ei,Carr:1974nx}.  Contrary to the astrophysical
processes for which only \acp{BH} heavier than $\mathcal{O}(1)$ solar
mass can be produced \cite{Rhoades:1974fn}, the small-mass \acp{BH}
could be also produced by the strong gravity inside the highly
compressed overdensities in the early Universe \cite{Carr:2005zd}. The
\ac{PBH} mass depends on the \ac{PBH} formation redshift $z_{f}$,
namely $M\simeq 30M_{\odot}[4\times10^{11}/(1+z_{f})]^{2}$
\cite{Sasaki:2016jop}, where $M_{\odot}$ is the solar mass
($=2\times 10^{33} {\rm g}$).  Since  inflation models
\cite{Starobinsky:1979ty,Starobinsky:1980te,Guth:1980zm,Linde:1981mu,Albrecht:1982wi,Sato:1980yn,
  Linde:1983gd} predict the properties of the primordial curvature
perturbations, which determine the mass and abundance of \acp{PBH}
(see
e.g. Refs.~\cite{Kohri:2007qn,Carr:2009jm,Sasaki:2018dmp,Byrnes:2018txb,Yoo:2018esr}
and references therein), our observational knowledge of the \acp{PBH}
is important to learn about the physics of the inflationary Universe.

Recently, it has been proposed that the energy-density fraction of
\acp{PBH} can be constrained by measuring the energy-density spectrum
of the \ac{SGWB}. First, the \ac{SGWB} can be produced from an
incoherent superposition of \acp{GW} emitted from all the coalescing
\ac{PBH} binaries. The null detection of such a \ac{SGWB} by the first
\ac{aLIGO} observing run \cite{TheLIGOScientific:2016dpb} has been
used to independently constrain $f_{\mathrm{PBH}}$
\cite{Wang:2016ana,Raidal:2017mfl,Mandic:2016lcn,Clesse:2016ajp}. For
example, Ref.~\cite{Wang:2016ana} obtained the tightest observational constraint on $f_{\mathrm{PBH}}$ in the mass range
$1-10^{2}M_{\odot}$, pushing the existing observational
constraints tighter by one order of magnitude. The possibility to
detect the \ac{SGWB} from \acp{PBH}, in particular from subsolar-mass
\acp{PBH}, by upcoming \ac{aLIGO} observing runs was also predicted
\cite{Wang:2016ana}. Second, the \ac{SGWB} is induced from the
enhanced primordial curvature perturbations~\cite{Mollerach:2003nq,
  Ananda:2006af, Baumann:2007zm, Assadullahi:2009jc}~\footnote{Based
  on the inflation model, primordial \acp{GW}
  \cite{Grishchuk:1974ny,Starobinsky:1979ty} are decoupled with
  primordial curvature perturbations at the first order. However, the
  induced \acp{GW} can be generated from primordial curvature
  perturbations at the second order. Whether or not the primordial
  \acp{GW} are detected in the future
  \cite{Creminelli:2015oda,Huang:2015gca,Huang:2017gmp,Cabass:2015jwe,Escudero:2015wba,Errard:2015cxa,Kamionkowski:2015yta,Zhao:2015sla,Guzzetti:2016mkm,Lasky:2015lej},
  the induced \acp{GW} could be sizable and even be larger than the
  primordial \acp{GW}, if the primordial curvature perturbations are
  significantly enhanced
  \cite{Ananda:2006af,PhysRevD.81.023517,Alabidi:2012ex,Mollerach:2003nq,Baumann:2007zm,Assadullahi:2009nf,Alabidi:2013wtp,Cai:2019amo,Cai:2018dig,Unal:2018yaa}.}. By
making use of the semi-analytic calculation of the induced \ac{GW}
spectrum \cite{Kohri:2018awv,Espinosa:2018eve}, the null detection of
such a \ac{SGWB} by a variety of \ac{GW} detectors has been used to
obtain constraints on the spectral amplitude of primordial curvature
perturbations
\cite{Inomata:2016rbd,Inomata:2018epa,Orlofsky:2016vbd}. The
constraints on the induced \ac{SGWB} can be recast as the constraints
on the abundance of \acp{PBH}, and vice versa
\cite{PhysRevD.81.023517,Bugaev:2010bb,Saito:2008jc,Saito:2009jt,Nakama:2016enz}.

In this paper, we focus on the small-mass \acp{PBH} for $10^{-8}M_{\odot}\leq M_{\mathrm{PBH}}\leq 1M_{\odot}$. Correspondingly, we calculate the energy-density fraction of the above two kinds of \acp{SGWB}, and report the expected constraints on the energy-density fraction of \acp{PBH} from the null detection of the \acp{SGWB} by several on-going and planned \ac{GW} experiments (see details in Ref.~\cite{Pitkin:2011yk}), which include \ac{SKA} \cite{Moore:2014lga}, \ac{LISA} \cite{Audley:2017drz, Cornish:2018dyw}, \ac{DECIGO} \cite{Sato:2017dkf} and B-\ac{DECIGO} \cite{Isoyama:2018rjb}, \ac{BBO} \cite{Harry:2006fi}, \ac{ET} \cite{Punturo:2010zz}, and \ac{aLIGO} design sensitivity \cite{TheLIGOScientific:2016wyq}. Although we focus on the mass range $10^{-8} M_\odot \leq M_{\text{PBH}} \leq 1 M_\odot$, the method of our analysis is equally applicable to the PBS masses outside of this range.  In this context, the authors of Refs.~\cite{Bartolo:2018evs, Bartolo:2018rku} studied the SGWB induced by the curvature perturbations associated with the PBHs of masses around $10^{-12} M_\odot$ as it may explain the whole abundance of the dark matter.  The authors of Ref.~\cite{Inomata:2018epa} obtained the constraints on the primordial curvature perturbations by studying the detectability of the curvature-induced SGWB in a wide frequency range corresponding to a wide PBH mass range. 

First, following Ref.~\cite{Wang:2016ana}, we evaluate the energy-density fraction of the \ac{SGWB} from binary \ac{PBH} coalescence, by assuming a monochromatic mass distribution of \acp{PBH}. This choice of the delta function is reasonable since the mass distribution of \acp{PBH} is insensitive to the details of the spectral shape of primordial curvature perturbations especially after taking into account coarse graining within the Hubble horizon and the effects of critical collapse \cite{Byrnes:2018txb}. In addition, the inflation scenario does not favor a significantly extended \ac{PBH} mass distribution \cite{Carr:2016drx}.
Second, following Ref.~\cite{Kohri:2018awv}, we evaluate the energy-density fraction of the induced \ac{SGWB}, by assuming a delta function for the power spectrum of primordial curvature perturbations. In principle, the spectrum of the induced \ac{SGWB} depends on the details of the spectral shape of primordial curvature perturbations. Recently, Ref.~\cite{Inomata:2018epa} found a spread of the \ac{SGWB} spectrum by studying a log-normal distribution for the power spectrum of primordial curvature perturbations. So the results obtained by this work can be regarded as the conservative one.\footnote{This is not that simple because the spectral index of the tails of the SGWB is also relevant as well as the width around the peak, and the spectral index depends on the shape of the curvature perturbations.  See the discussion around Eq.~(58) of Ref.~\cite{Espinosa:2018eve}.  Anyway, we focus on the delta function case for definiteness.
}  See also Ref.~\cite{Clesse:2018ogk} which discusses the effects of a
broad spectrum.

The rest of the paper is arranged as follows. In Sec.~\ref{sec:forpbh}, we briefly review the formation of \acp{PBH} in the early Universe, given the power spectrum of primordial curvature perturbations. In Sec.~\ref{sec:sgwbpbh}, we evaluate the energy-density fraction of the \ac{SGWB} from binary \ac{PBH} coalescence, and use it to obtain expected constraints on $f_{\mathrm{PBH}}$ from a variety of on-going and planned \ac{GW} detectors. In Sec.~\ref{sec:isgwb}, we evaluate the induced \ac{SGWB} from the enhanced primordial curvature perturbations, and obtain the expected constraints on the energy-density fraction of \acp{PBH} from \ac{SKA} and \ac{LISA}. The conclusions and discussions are given in Sec.~\ref{sec:con}.

%%%%%%%%%%%%%%%%%%%%%%%%%%%%%%%%%%%%%%%%%%%%%%%%%%%%%%%%%%%%%%%%%%%%%%
\section{Formation of primordial black holes} \label{sec:forpbh}
%%%%%%%%%%%%%%%%%%%%%%%%%%%%%%%%%%%%%%%%%%%%%%%%%%%%%%%%%%%%%%%%%%%%%%
\noindent 
Given the power spectrum of the primordial curvature perturbations, we can evaluate a probability of the \ac{PBH} production, the  mass function of \ac{PBH}s and the \ac{PBH} abundance \cite{Carr:2009jm,Sasaki:2018dmp}.
In this work we assume that the \acp{PBH} are formed in the early Universe which is \ac{RD}.
First of all, we need to estimate the wavenumber scale $k$ which is related with a given mass scale $M_{H}$ within the Hubble horizon at the time of horizon re-entry. 
According to Appendix~\ref{append:a}, it is represented by
\begin{align}\label{eq:k-m}
\frac{k}{k_\ast} = 7.49 \times 10^7 
\left( \frac{M_\odot}{M_{H}} \right)^{1/2} \left( \frac{g_{\ast,\rho} (T(M_{H}))}{106.75} \right)^{1/4} \left( \frac{g_{\ast,s} (T(M_{H}))}{106.75} \right)^{-1/3}\ ,
\end{align}
where 
$k_\ast=0.05\mathrm{Mpc}^{-1}$. 
Here we can numerically obtain the temperature at the formation $T(M_{H})$ by using Eq.~(\ref{eq:mht}).
The effective degrees of freedom of relativistic particles, i.e. $g_{\ast \rho}$ and $g_{\ast s}$, are precisely calculated for the Standard Model in Ref.~\cite{Saikawa:2018rcs}. Here we interpolate the tabulated data provided by the associated website \footnote{\url{http://member.ipmu.jp/satoshi.shirai/EOS2018}} to this reference.

The phenomena of critical collapse \cite{Yokoyama:1998xd,Carr:2016drx}
could describe the formation of \acp{PBH} with mass $M$ in the early
Universe, depending on the horizon mass $M_{H}$ and the amplitude of
density fluctuation $\delta$.  We have the following relation
\be\label{eq:mmh}
M=K M_{H}\(\delta-\delta_{c}\)^{\gamma}\ ,
\ee
where $K=3.3$, $\gamma=0.36$ and $\delta_{c}={0.45}$ are numerical
constants \footnote{Analytically it is estimated to be
  $\delta_{c}={0.41}$~\cite{Harada:2013epa}. In fact, however, they
  depend on the radial profile of the density
  perturbations.~\cite{Harada:2015yda}}.
The above equation can be inverted to express $\delta$ in terms of $M/M_{H}$, namely, $\delta=(M/(KM_{H}))^{1/\gamma}+\delta_{c}$, which is useful in the following calculations.

In the \ac{RD} Universe, the coarse grained density perturbation is given by
\be\label{eq:sigma}
\sigma^{2}(k)=\int_{-\infty}^{+\infty}d\ln q\ w^{2}(q/k)\(\frac{4}{9}\)^{2}\(\frac{q}{k}\)^{4}T^{2}(q,\tau=1/k)P_{\zeta}(q)\ ,
\ee
where $w(q/k)=\mathrm{exp}(-q^{2}/(2k^{2}))$ is a Gaussian window function, and $T(q,\tau)=3(\sin y-y\cos y)/y^{3}$ ($y\equiv q\tau/\sqrt{3}$) is a transfer function (see e.g.~Refs.~\cite{Young:2014ana,Ando:2018qdb} for details). 
We consider the power spectrum of primordial curvature perturbations $P_{\zeta}(k)$ to be a delta function of $\ln k$, i.e.,
\be\label{eq:pzeta}
P_{\zeta}(k)=A\delta(\ln k - \ln k_{0})\ ,
\ee
where $k_{0}$ is a given constant wavenumber, and $A$ is a dimensionless amplitude.
By substituting Eq.~(\ref{eq:pzeta}) into Eq.~(\ref{eq:sigma}), we obtain
\be\label{eq:sigmak}
\sigma^{2}(k)=16A e^{-1/x^{2}}\[\cos^{2}\(\frac{1}{\sqrt{3}x}\)+x\(3x\sin^{2}\(\frac{1}{\sqrt{3}x}\)-\sqrt{3}\sin\(\frac{2}{\sqrt{3}x}\)\)\]\ ,
\ee
where $x \equiv k/k_{0}$ is a dimensionless quantity.
We show $\sigma^{2}(k)/A$ versus $k/k_{0}$ in a figure at the end of Appendix~\ref{append:a}.

To convert $\sigma(k)$ to the mass function of \acp{PBH}, by making
use of the Press-Schechter formalism \cite{Press:1973iz}, we calculate
the probability of the \ac{PBH} production, i.e.,
\be\label{eq:beta}
\beta_{M_H} =\int_{\delta_{c}}^{\infty}\frac{M}{M_{H}}\mathcal{P}_{M_H}(\delta(M))d\delta(M)
=\int_{-\infty}^{\infty}\frac{M}{M_{H}}\mathcal{P}_{M_H}(\delta(M))\frac{d\delta (M)}{d\ln M}d\ln M
\equiv \int_{-\infty}^\infty \widetilde{\beta}_{M_H}(M) d \ln M \ ,
\ee
which accounts for the fraction of the Hubble volumes that collapse
into \acp{PBH} when the horizon mass is $M_H$.
$\widetilde{\beta}_{M_H}(M)$ is the distribution of the (logarithmic)
masses of \acp{PBH} resulting after the critical collapse.  Here
$\mathcal{P}_{M_H}(\delta)$ denotes a Gaussian probability
distribution of primordial density perturbations at the given horizon
scale corresponding to $M_H$.  It is represented by
\be
\mathcal{P}_{M_H}(\delta(M))=\frac{1}{\sqrt{2\pi\sigma^{2}(k(M_{H}))}}\mathrm{exp}\(-\frac{\delta^{2}(M)}{2\sigma^{2}(k(M_{H}))}\)\ ,
\ee
where $\sigma(k(M_{H}))$ is computed by making use of
Eq.~(\ref{eq:sigmak}), and $k(M_{H})$ is given by Eq.~(\ref{eq:k-m}).
The explicit form of $\widetilde{\beta}_{M_H}(M)$ is written to be~\cite{Niemeyer:1997mt}
\begin{align}
\widetilde{\beta}_{M_H}(M) = \frac{K}{\sqrt{2\pi}\gamma \sigma(k(M_H))} \left( \frac{M}{K M_H} \right)^{1+\frac{1}{\gamma}} \exp \left( -  \frac{1}{2 \sigma^2 (k(M_H))} \left( \delta_c + \left( \frac{M}{K M_H} \right)^{\frac{1}{\gamma}} \right)^2\right).
\end{align}
The mass function of \acp{PBH} is defined as
$f(M)=\frac{1}{\Omega_{\mathrm{CDM}}}\frac{d\Omega_{\mathrm{PBH}}}{d\ln
  M}$,
and the abundance of \acp{PBH} in \ac{CDM} is given by
$f_{\mathrm{PBH}}=\int f(M) \text{d} \ln (M/M_\odot)$.
We obtain the mass function of \acp{PBH} as follows (see
e.g. Ref.~\cite{Byrnes:2018txb})
\be
&&f(M)=\frac{\Omega_{\mathrm{m}}}{\Omega_{\mathrm{CDM}}}\int_{-\infty}^{\infty}\(\frac{g_{\ast,\rho}(T(M_{H}))}{g_{\ast,\rho}(T_{\mathrm{eq}})}\frac{g_{\ast,s}(T_{\mathrm{eq}})}{g_{\ast,s}(T(M_{H}))}\frac{T(M_{H})}{T_{\mathrm{eq}}}\) \widetilde{\beta}_{M_H} (M)d\ln M_{H}\ ,
\ee
where $T_{\mathrm{eq}}$ is the temperature of the Universe at the epoch of matter-radiation equality.
\begin{figure}[tbh]
 \centering
\includegraphics[width=0.64 \columnwidth]{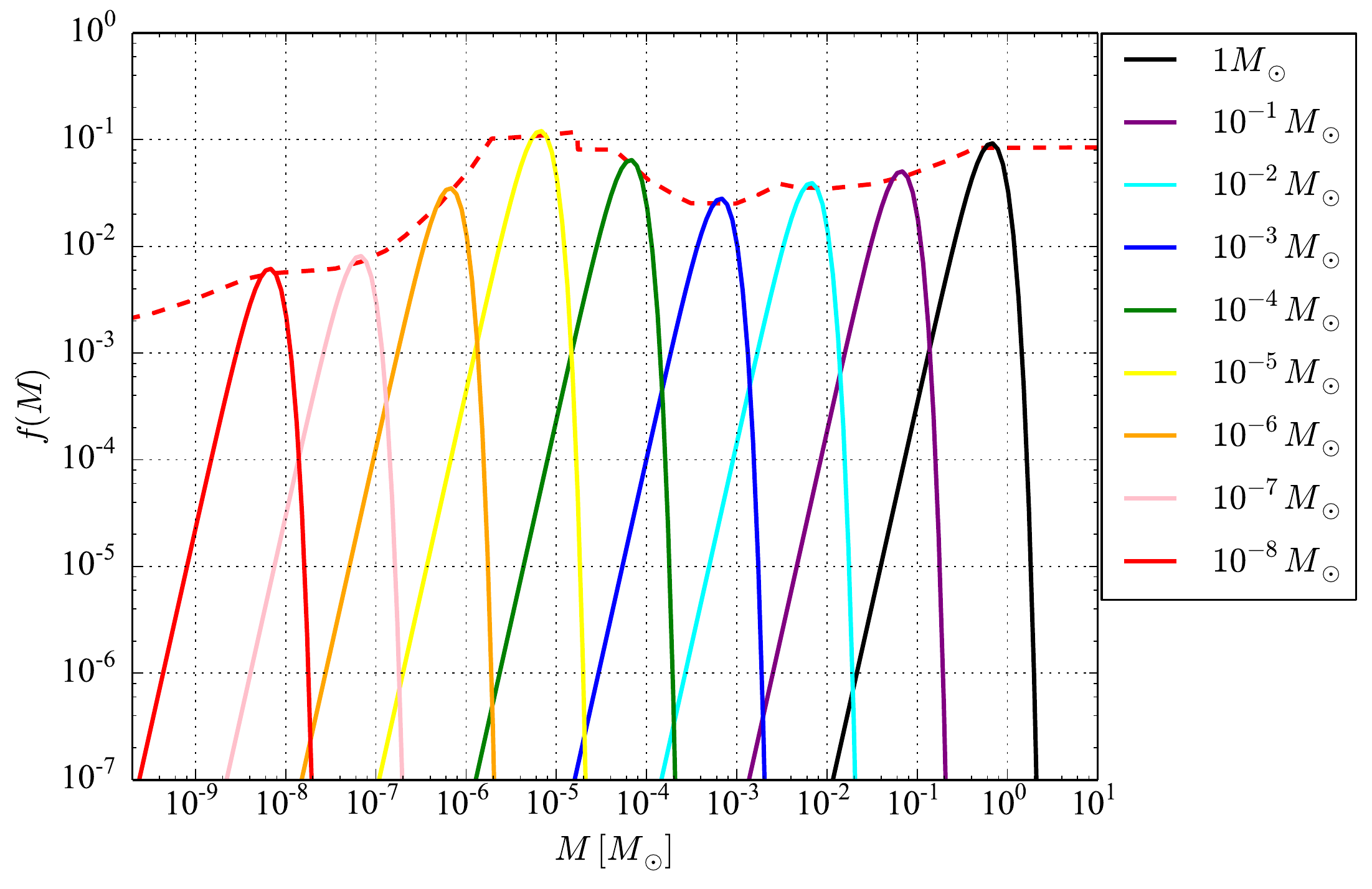}
\caption{Mass functions $f(M)$ originated from the delta function power spectrum $P_\zeta (k)$ (colored solid).  From right to left, based on Eq.~(\ref{eq:k-m}), the value of $k_0$ is chosen so that the corresponding $M_{H}$ is $\log_{10} (M_{H}/M_\odot)= 0,-1,-2,-3,-4,-5,-6,-7,-8$.  The normalization $A$ is chosen so that $f_{\mathrm{PBH}}$ is equal to the upper bound on $f(M)$ with the aforementioned values of $M_{H}$.
To be specific, we take $10^{2}A$ as 5.8898, 5.4658, 5.1116, 4.8011, 4.6799, 4.6490, 4.3002, 3.8740 and 3.6813, respectively.
The existing  observational constraints (HSC~\cite{Niikura:2017zjd} (green dotted), OGLE~\cite{Niikura:2019kqi} (blue dotted), EROS/MACHO~\cite{Tisserand:2006zx, Allsman:2000kg} (cyan dotted), caustic crossing \cite{Oguri:2017ock} (purple dotted) and their combination (red dashed)) are plotted for comparison.
  }
 \label{fig:pbhmassdist}
\end{figure}
In Fig.~\ref{fig:pbhmassdist}, we depict several examples (colored
solid) for the mass function $f(M)$ which is originated from
$P_\zeta(k)$ in
Eq.~(\ref{eq:pzeta}). 
To be specific, we choose a horizon mass to be
$M_{H}=10^{-i}M_{\odot}~(i=0,1,2,...,8)$, each of which determines the
value of its own $k_{0} = k_{0}(M_{H})$.  Here $(10^{2}A)$ is 5.8898,
5.4658, 5.1116, 4.8011, 4.6799, 4.6490, 4.3002, 3.8740 and 3.6813,
respectively, so that $f_{\mathrm{PBH}}$ equals the upper limit on
$f(M)$ with the aforementioned values of $M_{H}$.  For comparison, we
plot the existing observational constraint (red dashed) on the \ac{PBH} mass
function. The constraint used here arises from the microlensing
observations of Subaru/HSC \cite{Niikura:2017zjd}, OGLE
\cite{Niikura:2019kqi}, EROS-2 \cite{Tisserand:2006zx}, MACHO
\cite{Allsman:2000kg}, and the caustic crossing \cite{Oguri:2017ock}.

%%%%%%%%%%%%%%%%%%%%%%%%%%%%%%%%%%%%%%%%%%%%%%%%%%%%%%%%%%%%%%%%%%%%%%
\section{Stochastic gravitational-wave background due to binary primordial black hole mergers} 
%%%%%%%%%%%%%%%%%%%%%%%%%%%%%%%%%%%%%%%%%%%%%%%%%%%%%%%%%%%%%%%%%%%%%%
\label{sec:sgwbpbh}
\noindent 
Two different mechanisms have been proposed to form binaries from the
\acp{PBH}. One scenario assumes that two \acp{PBH} could form a binary
due to the energy loss via gravitational radiation when they pass by
each other accidentally in the late Universe
\cite{Bird:2016dcv,Clesse:2016vqa}. The other one assumes that two
nearby \acp{PBH} form a binary due to the tidal force from a third
neighboring \ac{PBH} in the early Universe
\cite{Nakamura:1997sm,Ioka:1998nz,Sasaki:2016jop}. Both scenarios are
capable of explaining the merger rates of \acp{BBH} reported by
\ac{aLIGO}. However, the first one requires the \acp{PBH} to
contribute most of the \ac{CDM}, which is disfavored by various
observational constraints in the relevant mass range.  On the other
hand, the second one is still allowed. In this work, we thus adopt the
formation scenario \footnote{In this section, we use the revised
  formalism in Ref.~\cite{Sasaki:2018dmp}, instead of the original one
  in Ref.~\cite{Sasaki:2016jop}. } of \ac{PBH} binaries proposed in
Ref.~\cite{Nakamura:1997sm} and revisited by
Refs.~\cite{Ioka:1998nz,Sasaki:2016jop,Ali-Haimoud:2017rtz,Eroshenko:2016hmn,Hayasaki:2009ug,Ballesteros:2018swv,Chen:2018czv,Raidal:2018bbj,Liu:2018ess,Abbott:2018oah,Magee:2018opb}. In
Appendix~\ref{append:mr} we show a brief summary of the formalism for
such a scenario.

We calculate the \ac{SGWB} spectrum produced from the coalescing
\ac{PBH} binaries.  In general, the dimensionless energy-density
spectrum of the \ac{SGWB} is defined as
$\Omega_{\mathrm{GW}}=\rho_{\mathrm{c}}^{-1}d\rho_{\mathrm{GW}}/d\ln\nu$,
where $\rho_{\mathrm{GW}}$ is the \ac{GW} energy density, and $\nu$ is
the \ac{GW} frequency \cite{Allen:1997ad}.  Knowing the merger rate of
\ac{PBH} binaries in Eq.~(\ref{merger rate}), according to
Ref.~\cite{Wang:2016ana}, we can compute the \ac{SGWB} energy-density
spectrum within the frequency interval $(\nu,\nu+d\nu)$. It is given
by
\begin{align}
\Omega_{\textrm{GW}}(\nu) = \frac{\nu}{\rho_\textrm{c}} \int_{0}^{\frac{\nu_{\mathrm{cut}}}{\nu}-1} \frac{R_\textrm{PBH}(z)}{(1+z)H(z)}  \frac{dE_\textrm{GW}}{d\nu_\textrm{s}}(\nu_\textrm{s}) dz\ ,
\label{GW spectrum}
\end{align}
where $\frac{dE_\textrm{GW}}{d\nu_\textrm{s}}(\nu_\textrm{s})$ is the \ac{GW} energy spectrum of a \ac{BBH} coalescence (see details in Refs.~\cite{Ajith:2007kx,Ajith:2009bn}, or a brief summary in Appendix~\ref{append:d}), $\nu_\textrm{s}$ is the frequency in the source frame and is related to the observed frequency $\nu$ through $\nu_\textrm{s}=(1+z)\nu$, and $\nu_{\mathrm{cut}}$ is the cutoff frequency for a given \ac{BBH} system.

\begin{figure}[tbh]
 \centering
\includegraphics[width=0.64 \columnwidth]{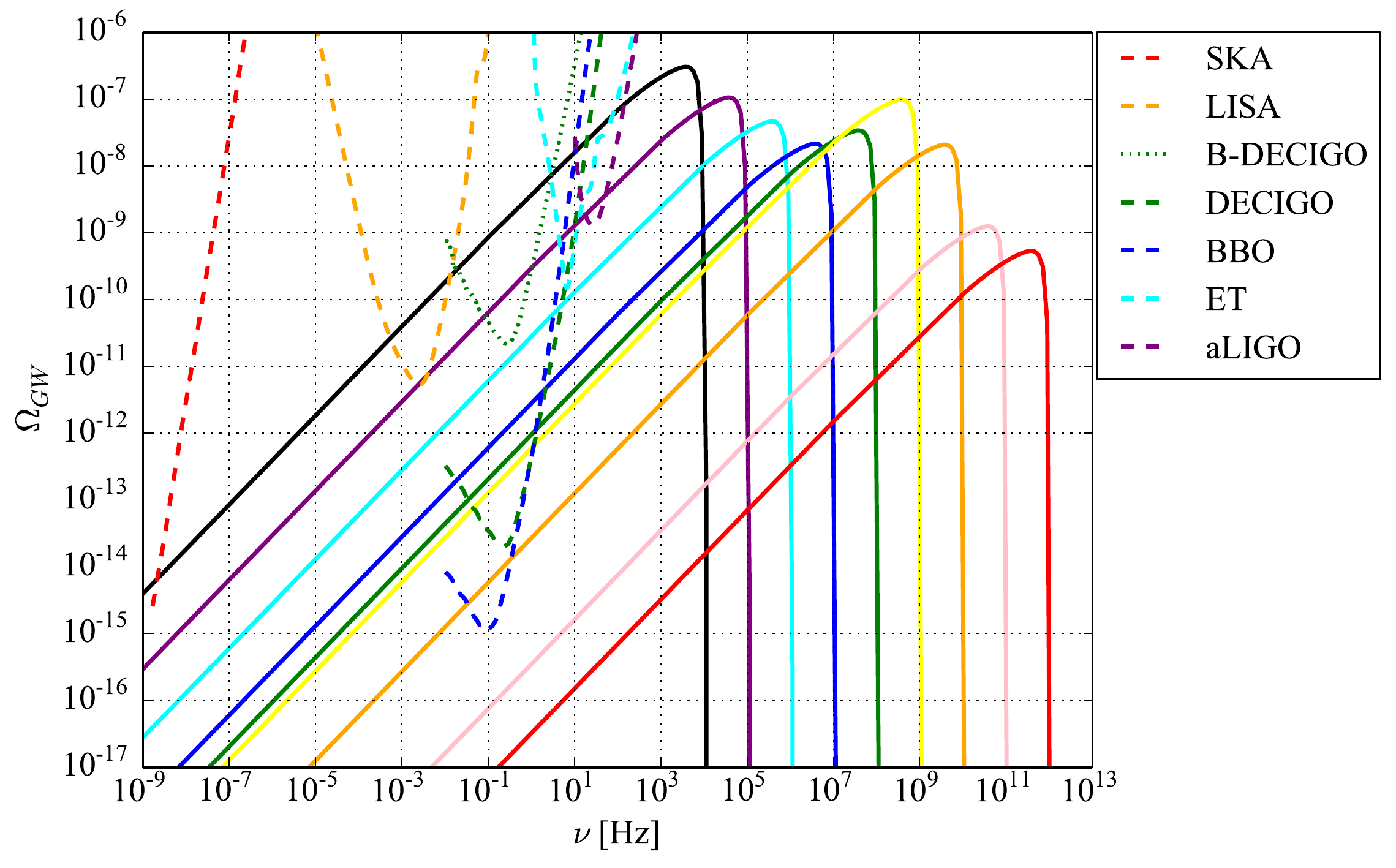}
  \caption{Energy-density spectrum (colored solid) of the \ac{SGWB} due to binary \ac{PBH} coalescence which is just allowed by the existing  observational constraints on the \ac{PBH} abundance. The \ac{SGWB} spectra with the cutoff frequencies from right to left correspond to the peaks from left to right in Fig.~\ref{fig:pbhmassdist} (same colors). 
The sensitivity curves (colored dashed/dotted) of the \ac{GW} detectors are also plotted for comparison. 
  }
 \label{fig:sgwb-pbh}
\end{figure}
\begin{figure}[tbh]
 \centering
\includegraphics[width=0.53 \columnwidth]{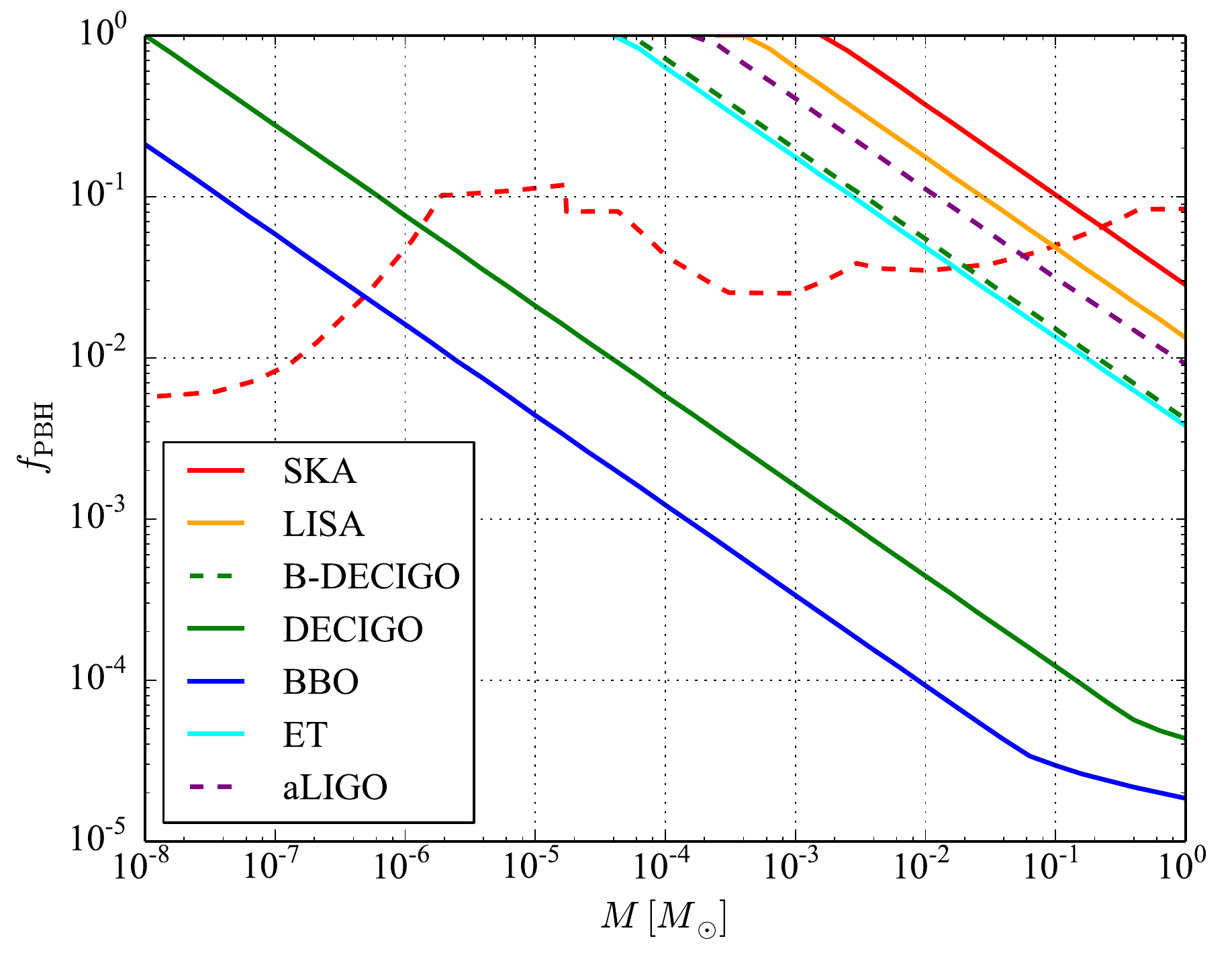}
  \caption{Expected constraints on the \ac{PBH} abundance from the null detection of the \ac{SGWB} by \ac{LISA} (orange solid), B-\ac{DECIGO} (green dashed), \ac{DECIGO} (green solid), \ac{BBO} (blue solid), \ac{ET} (cyan solid), and \ac{aLIGO} (purple dashed). The present existing constraint (red dashed) is plotted for comparison.}
 \label{fig:expectedconstraints}
\end{figure}
For the \ac{PBH} binaries for the component masses
$10^{-i}M_{\odot}~(i=0,1,2,...,8)$, which correspond to the examples
of the \ac{PBH} mass function in Fig.~\ref{fig:pbhmassdist}, we plot
the corresponding energy-density fractions of the \ac{SGWB} due to
binary \ac{PBH} coalescence at the existing observational constraints on the
\ac{PBH} abundance in
Fig.~\ref{fig:sgwb-pbh}. 
The color coding is the same as that in Fig.~\ref{fig:pbhmassdist}.
For comparison, we depict the sensitivity curves \footnote{Sometimes,
  only the amplitude spectral density $S_{n}(f)$ is shown for a given
  gravitational-wave detector. We have
  $\sqrt{S_{n}(f)}=h_{n}(f)f^{-1/2}$, which has the unit of
  $\mathrm{Hz}^{-1/2}$, and $h_{n}$ is the noise amplitude. The
  sensitivity to the \ac{SGWB} energy density is related with
  $S_{n}(f)$ by
  $\Omega_{\mathrm{GW},n}(f)=3.132\times10^{35}h^{-2}(f/\mathrm{Hz})^{3}(\sqrt{S_{n}(f)}/\mathrm{Hz}^{-1/2})^{2}$
  \cite{Moore:2014lga,Kikuta:2014eja}. The reduced Hubble constant is
  $h=0.678$ in this paper.}  of several \ac{GW} experiments (colored
dashed/dotted curves), which include pulsar timing array (\ac{SKA}
\cite{Moore:2014lga}), space-based \ac{GW} interferometers (\ac{LISA}
\cite{Cornish:2018dyw},
\ac{DECIGO} \cite{Sato:2017dkf} and B-\ac{DECIGO}
\cite{Isoyama:2018rjb}, \ac{BBO} \cite{Harry:2006fi}),
third-generation ground-based \ac{GW} interferometer (\ac{ET}
\cite{Punturo:2010zz}) and second-generation ground-based \ac{GW}
interferometer (\ac{aLIGO} \cite{TheLIGOScientific:2016wyq}).
If the spectrum predicted in a model intersects the sensitivity curve
of a given experiment, the expected \ac{SNR} is equal to or greater
than unity, which means a possible detection of such a spectrum by
this experiment.

Null detection of the \ac{SGWB} by the given future or on-going
\ac{GW} experiment can place an upper bound on the magnitude of the
energy-density fraction of the \ac{SGWB} at a given frequency band,
and can be further recast to constrain the maximum \ac{PBH}
abundance. From Fig.~\ref{fig:sgwb-pbh}, all the \ac{GW} experiments 
have possible contributions to improve
the existing observational constraints on the \ac{PBH} abundance, since their
sensitivity curves intersect some spectra. Therefore, by regarding the
sensitivity curves of all these experiments as upper bounds on the
\ac{SGWB} spectrum, we evaluate the expected upper limits on the
\ac{PBH} abundance from these experiments. We depict our results in
Fig.~\ref{fig:expectedconstraints}.

Our results are as follows. \ac{SKA}, \ac{LISA} and \ac{aLIGO} will give us
relatively weak constraints in future. It is notable that this expected limit
from aLIGO is surely stronger than the current one which was reported
recently by Ref.~\cite{Abbott:2018oah}. Both \ac{ET} and
B-\ac{DECIGO} also have similar constraints on the abundance. All the
above four experiments are expected to improve the existing observational
constraints on the subsolar-mass \acp{PBH}. However, both \ac{DECIGO}
and \ac{BBO} are expected to significantly improve the existing
constraints over the mass range
$\mathcal{O}(10^{-6})\leq M/M_{\odot} \leq \mathcal{O}(10^{0})$.

%%%%%%%%%%%%%%%%%%%%%%%%%%%%%%%%%%%%%%%%%%%%%%%%%%%%%%%%%%%%%%%%%%%%%%
\section{Stochastic gravitational-wave background induced by primordial curvature perturbations}
%%%%%%%%%%%%%%%%%%%%%%%%%%%%%%%%%%%%%%%%%%%%%%%%%%%%%%%%%%%%%%%%%%%%%%
\label{sec:isgwb}
\noindent 
The \ac{SGWB} can be also induced by the enhanced primordial curvature
perturbations via the scalar-tensor mode coupling in the second-order
perturbation theory~\cite{Ananda:2006af}. In Appendix~\ref{append:b},
we show a brief summary of the evaluations of the induced \ac{SGWB}
spectrum. For details, see Ref.~\cite{Kohri:2018awv} and references
therein.  In the following, we will use Appendix~\ref{append:b} to
calculate the energy-density spectrum of the induced \ac{SGWB}, given
the form of $P_{\zeta}(k)$ in Eq.~(\ref{eq:pzeta}).
We consider the minimal case in which the statistics of the curvature
perturbations is Gaussian 
\footnote{The statistical properties of curvature perturbations can modify the relation between the amount of induced \ac{SGWB} and the \ac{PBH} abundance. Even the curvature perturbations are completely Gaussian, the density contrasts are non-Gaussian due to the nonlinear nature of the gravity, as shown recently by Refs.~\cite{Yoo:2018esr, Kawasaki:2019mbl,DeLuca:2019qsy,Young:2019yug}.} 
and neglect the time evolution of the mass
function of PBHs due to accretion, but generalizations can be found in
Ref.~\cite{Garcia-Bellido:2017aan}.

According to Eqs.~(\ref{eq:isgwb})--(\ref{eq:i2}), we obtain the dimensionless energy-density spectrum of the induced \ac{SGWB} as 
\be\label{eq:isgwbppp}
\Omega_{\mathrm{IGW}}\left(\nu=\frac{k}{2\pi}\right)&=&\Omega_{\mathrm{r},0}\(\frac{g_{\ast}(T(k))}{g_{\ast}(T_{\mathrm{eq}})}\)\(\frac{g_{\ast,s}(T(k))}{g_{\ast,s}(T_{\mathrm{eq}})}\)^{-4/3}
\times\frac{3A^{2}}{64}\(\frac{4-\tilde{k}^{2}}{4}\)^{2}\tilde{k}^{2}\(3\tilde{k}^{2}-2\)^{2}\nonumber\\
&&\times\[\pi^{2}\(3\tilde{k}^{2}-2\)^{2}\Theta(2-\sqrt{3}\tilde{k})+\(4+\(3\tilde{k}^{2}-2\)\ln\bigg|1-\frac{4}{3\tilde{k}^{2}}\bigg|\)^{2}\]\Theta(2-\tilde{k})\ ,
\ee
where $\nu=k/2\pi $ denotes the frequency of \ac{GW}, and the dimensionless wavenumber $\tilde{k}=k/k_{0}$ is introduced for simplicity.
Based on Appendix~\ref{append:a}, cosmic temperature $T$ can be numerically related with $M_{H}$ and $k$, and then with $\nu$.
Here $\Theta(x)$ denotes the Heaviside theta function with variable $x$.

\begin{figure}[tbh]
 \centering
\includegraphics[width=0.53 \columnwidth]{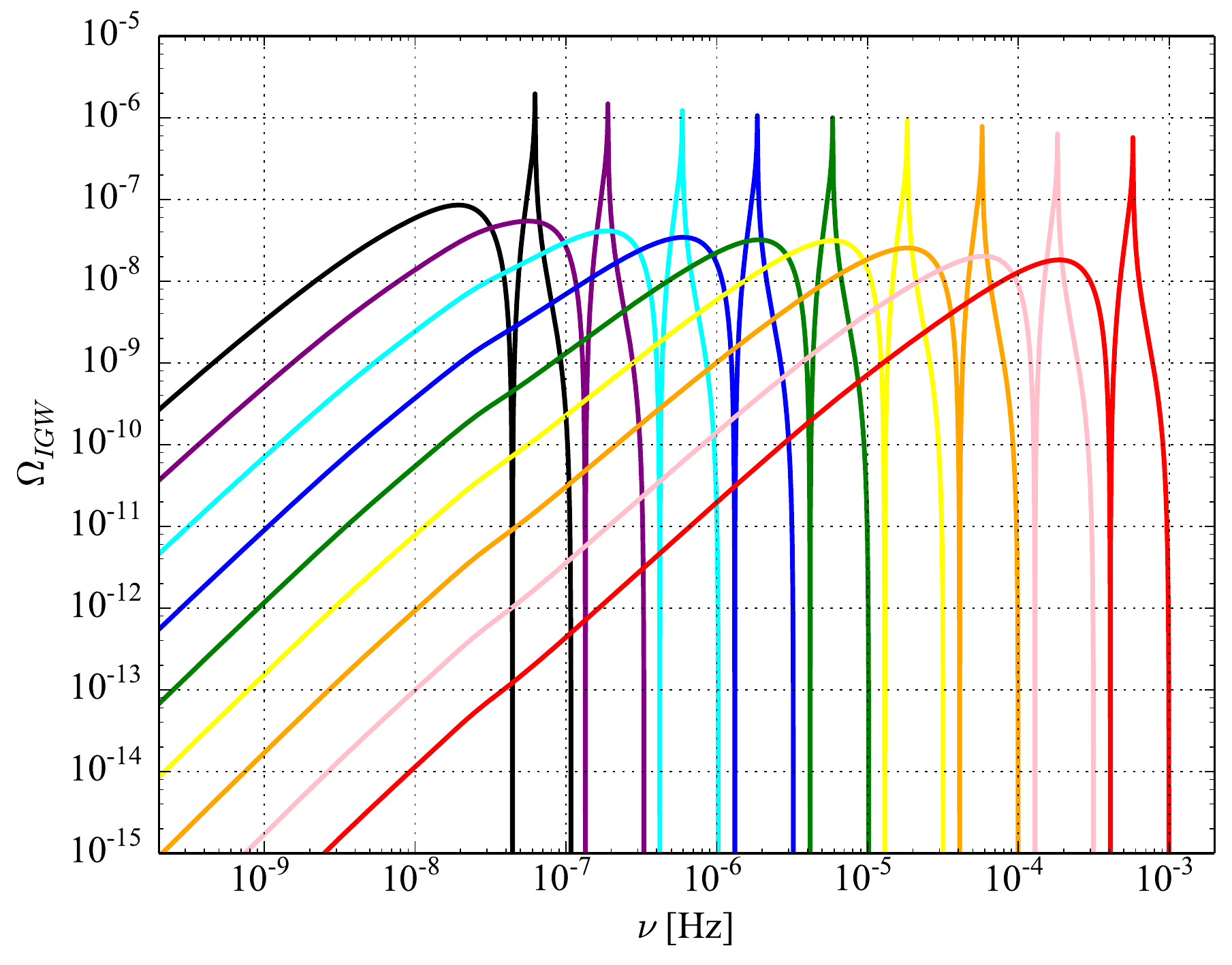}
  \caption{Energy-density spectrum of the \ac{SGWB} nonlinearly induced by the primordial curvature perturbations. The \ac{SGWB} spectra (colored solid) with the peaks from right to left correspond to the mass functions with the peaks from left to right in Fig.~\ref{fig:pbhmassdist} (same colors).}
 \label{fig:isgwb-scalar}
\end{figure}
\begin{figure}[tbh]
 \centering
\includegraphics[width=0.53 \columnwidth]{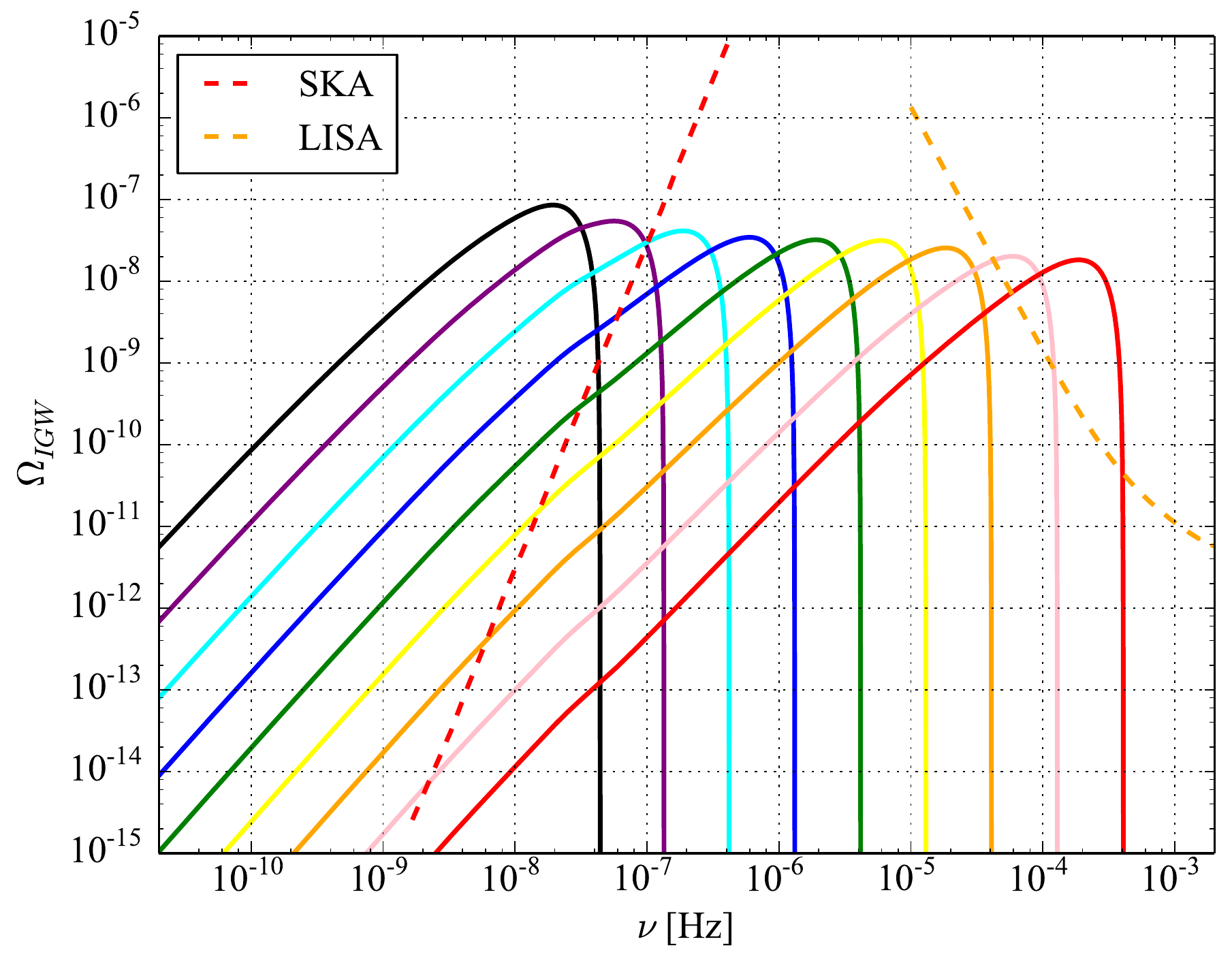}
\caption{Similarly to Fig.~\ref{fig:isgwb-scalar}, we plot the energy-density spectrum of the induced \ac{SGWB} (colored solid), but the right-handed peak is conservatively dropped. We depict the sensitivity curves of SKA (red dashed) and LISA (orange dashed) for comparison.}
 \label{fig:isgwb-scalar-constraints}
\end{figure}
\begin{figure}[tbh]
 \centering
\includegraphics[width=0.53 \columnwidth]{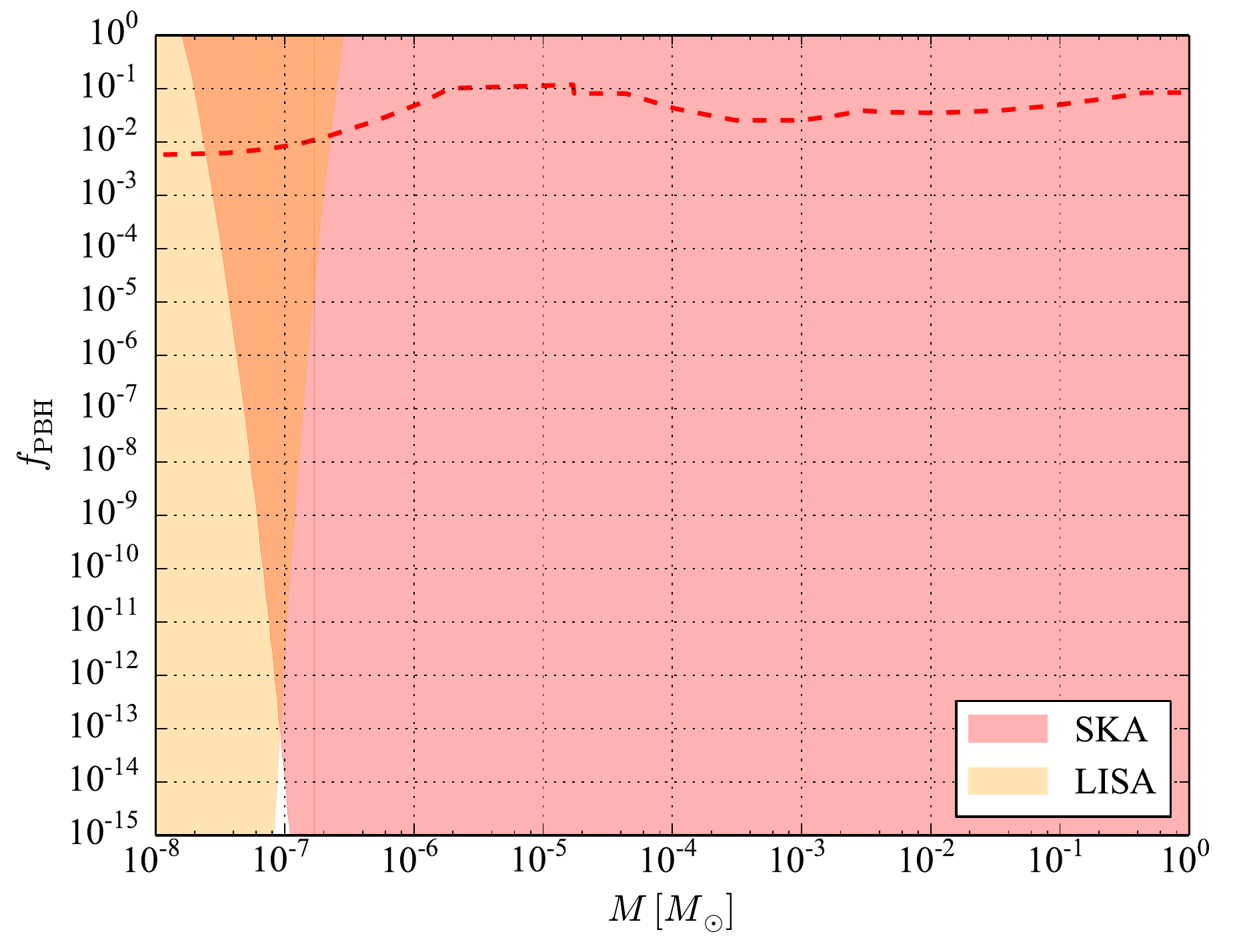}
  \caption{Expected constraints on the \ac{PBH} abundance versus the \ac{PBH} mass from the null detection of the induced \ac{SGWB} by \ac{SKA} (red shaded) and \ac{LISA} (orange shaded). 
    The existing observational constraint (red dashed) is also
    plotted for comparison.}
 \label{fig:expectedconstraints-1}
\end{figure}

Similarly to Fig.~\ref{fig:sgwb-pbh}, we plot the energy-density
fractions of the induced \ac{SGWB} due to the enhanced primordial
curvature perturbations in Fig.~\ref{fig:isgwb-scalar}. Both $A$ and
$k_{0}$ are chosen as those in Sec.~\ref{sec:forpbh}. The same
color coding is used as in Fig.~\ref{fig:pbhmassdist}. The double-peak
structures arise from the property of the delta function for
$P_{\zeta}(k)$ in Eq.~(\ref{eq:pzeta}). For a broader distribution for
$P_{\zeta}(k)$, e.g., a log-normal distribution in
Ref.~\cite{Inomata:2018epa}, one could find a spread of the \ac{SGWB}
spectrum. Therefore, our discussions in the next two paragraphs could
be regarded as conservative.

In Fig.~\ref{fig:isgwb-scalar-constraints}, besides the energy-density
spectra of the induced \ac{SGWB} (colored solid, same as
Fig.~\ref{fig:isgwb-scalar}), we depict the sensitivity curves of
\ac{SKA} \cite{Moore:2014lga} (red dashed) and \ac{LISA}
\cite{Cornish:2018dyw} (orange dashed) for comparison. For a given
spectrum of the induced \ac{SGWB}, we conservatively drop the
right-handed peak since such a spiky structure
exists only for source spectra with a tiny width. Similarly to the
discussions in last section, if a model-predicting spectrum intersects
the sensitivity curve of a given experiment, it is possible to measure
such a spectrum by this experiment. In such a case, both \ac{SKA} and
\ac{LISA} are expected to exclude most of the parameter space, or
equivalently improve the existing observational constraints on the
\ac{PBH} abundance significantly. 

Assuming the null detection of the induced \ac{SGWB} from the enhanced primordial curvature perturbations, similarly to Fig.~\ref{fig:expectedconstraints}, we plot the expected constraints on the \ac{PBH} abundance from \ac{SKA} (red shaded) and \ac{LISA} (orange shaded) in Fig.~\ref{fig:expectedconstraints-1}. The shaded regions mean the excluded parts of the parameter space by these experiments. 
In fact, here we first obtain the constraints on $A$ from the induced \ac{SGWB}, and then recast them as the upper limits on $f_{\text{PBH}}$ according to the formulae in Sec.~\ref{sec:forpbh}.

\begin{figure}[tbh]
 \centering
\includegraphics[width=0.64 \columnwidth]{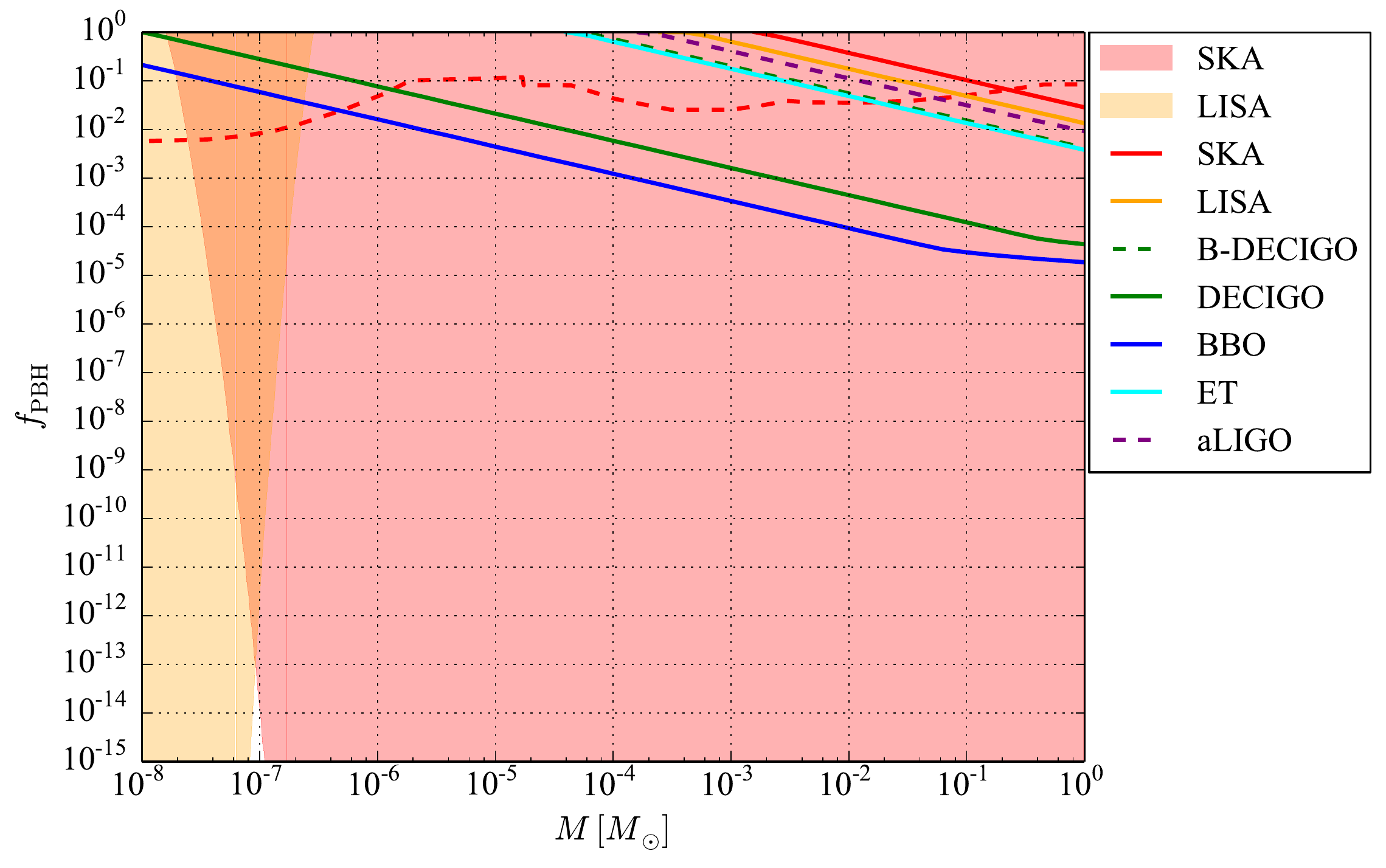}
  \caption{Expected constraints on the \ac{PBH} abundance versus the \ac{PBH} mass from the null detection of the two kinds of \acp{SGWB}. The existing observational constraint (red dashed) is plotted for comparison.}
 \label{fig:expectedconstraints-2}
\end{figure}
Finally, we can combine the results in
Fig.~\ref{fig:expectedconstraints} with
Fig.~\ref{fig:expectedconstraints-1} to obtain
Fig.~\ref{fig:expectedconstraints-2}.  Generally speaking, the slopes
of the upper bounds (i.e., boundaries of shaded regions) from the
induced \ac{SGWB} are significantly sharper than those (i.e., colored
curves) from the \ac{SGWB} due to the coalescing \ac{PBH}
binaries. This property can be easily understood as follows. On the
one hand, we directly constrained the magnitude of $f_{\mathrm{PBH}}$
by calculating the \ac{SGWB} from the coalescing \ac{PBH}
binaries. The detection of such a \ac{SGWB} requires a significant
amount of \ac{PBH} binaries in the Universe. This implies that the
\ac{GW} detectors can probe the enhanced primordial curvature
perturbations only if $A\sim\mathcal{O}(0.1)$. When
$A\ll \mathcal{O}(0.1)$, there would be so few \acp{PBH} in the
Universe that the thresholds of \ac{GW} detectors are not
triggered. On the other hand, by detecting the induced \ac{SGWB}, we
directly obtained the constraints on $A$, which were recast as the
indirect constraints on $f_{\mathrm{PBH}}$. In fact, by detecting the
induced \ac{SGWB}, the \ac{GW} detectors can probe the primordial
curvature perturbations of arbitrary amplitudes within their
sensitivities. Since the induced \ac{SGWB} spectrum is proportional to
$A^{2}$ while $f_{\mathrm{PBH}}$ is exponentially sensitive to $A$, we
obtained the sharper slopes for the upper bounds from the induced
\ac{SGWB} than those from the \ac{SGWB} due to coalescing \ac{PBH}
binaries in Fig.~\ref{fig:expectedconstraints-2}.  Thus, the
constraints on $f_{\text{PBH}}$ from the \ac{SGWB} induced by the
curvature perturbations are stronger than those from the \ac{SGWB}
whose origin is merger events except for a narrow gap around 
$10^{-7} M_{\odot}$ corresponding to the relatively weak observational sensitivity around about $10^{-6}$Hz.
Nevertheless, both types of the \ac{SGWB} are complementary and useful
for the consistency check of the PBH hypothesis since those two types
of the \ac{SGWB} have their own individual features in the spectra and
are probed by different observations which are supposed to measure GWs
at different frequency bands.

%%%%%%%%%%%%%%%%%%%%%%%%%%%%%%%%%%%%%%%%%%%%%%%%%%%%%%%%%%%%%%%%%%%%%%
\section{Conclusions} 
%%%%%%%%%%%%%%%%%%%%%%%%%%%%%%%%%%%%%%%%%%%%%%%%%%%%%%%%%%%%%%%%%%%%%%
\label{sec:con}
\noindent 
It has been known that \acp{PBH} can form binaries in the early Universe, and a \ac{PBH} binary can merge to a new heavier \ac{BH} due to the energy loss via gravitational radiation. Based on Ref.~\cite{Sasaki:2016jop}, the merger rate of \ac{PBH} binaries depends on the abundance and mass of \acp{PBH}. Given the existing constraints on the mass function of \acp{PBH}, following Ref.~\cite{Wang:2016ana}, we have evaluated the energy-density spectra of \acp{SGWB} which arise from coalescences of \ac{PBH} binaries with component masses $10^{-i}M_\odot~(i=0,1,2,...,8)$. From Fig.~\ref{fig:sgwb-pbh}, we found that some of them intersect the sensitivity curves of several future \ac{GW} experiments. This means that the existing limits can be improved by these experiments in the future if these experiments do not detect the \ac{SGWB}. By making use of these sensitivity curves as upper limits on the \ac{SGWB} energy-density fraction, we have evaluated the expected upper limits on the abundance of \acp{PBH}, and shown our results in Fig.~\ref{fig:expectedconstraints}. In particular, both DECIGO and BBO are expected to significantly improve the existing limits over the mass range $10^{-6}M_\odot-10^{0}M_\odot$.

The generation of \acp{PBH} in the early Universe requires large amplitudes of the primordial curvature perturbations, which can always induce the \ac{SGWB}. By taking into account the existing constraints on the mass function of \acp{PBH} and making use of the semi-analytic formula in Ref.~\cite{Kohri:2018awv}, we have calculated the energy-density spectrum of the induced \ac{SGWB}, and shown our results in Fig.~\ref{fig:isgwb-scalar}. We find several intersections between the induced \ac{SGWB} spectra and the sensitivity curves of \ac{SKA} and \ac{LISA} in Fig.~\ref{fig:isgwb-scalar-constraints}. This implies that these experiments can improve the existing upper limits on the mass function of \acp{PBH} in the future if they claim the null detection of the induced \ac{SGWB} energy-density fraction. In this case, the shaded regions in Fig.~\ref{fig:expectedconstraints-1} will be excluded by \ac{SKA} and \ac{LISA}, respectively.

Finally, by combining
Fig.~\ref{fig:expectedconstraints} with
Fig.~\ref{fig:expectedconstraints-1} to obtain
Fig.~\ref{fig:expectedconstraints-2},
we found stronger constraints on $f_{\mathrm{PBH}}$ from the \ac{SGWB} induced by curvature perturbations than those from the \ac{SGWB} due to coalescing events, except for a narrow gap around $10^{-7} M_{\odot}$.
However, both types of the \ac{SGWB} are complementary and useful for the consistency check of the \ac{PBH} hypothesis.

\vspace{0.1cm}
\begin{acknowledgements}
\noindent
This work is supported in part by the JSPS Research Fellowship for
Young Scientists (TT) and JSPS KAKENHI Grants No.~JP17H01131 (SW and
KK) and No.~JP17J00731 (TT), and MEXT KAKENHI Grants No.~JP15H05889
(KK), and No.~JP18H04594 (KK).
\end{acknowledgements}
\vspace{0.1cm}

%\newpage
\appendix
%%%%%%%%%%%%%%%%%%%%%%%%%%%%%%%%%%%%%%%%%%%%%%%%%%%%%%%%%%%%%%%%%%%%%%
\section{Relation between $k$ and $M_{H}$ in the radiation dominated Universe} 
%%%%%%%%%%%%%%%%%%%%%%%%%%%%%%%%%%%%%%%%%%%%%%%%%%%%%%%%%%%%%%%%%%%%%%
\label{append:a}
\noindent
During the radiation dominated (\ac{RD}) era of the Universe, the relation between the wavenumber $k$ and the horizon mass $M_{H}$ is obtained as follows. 
By definition, we have 
\begin{align}\label{eq:defkah}
k = a(M_{H}) H(M_{H})\ .
\end{align}
The value of the scale factor $a(M_{H})$, when the mode corresponding to $M_{H}$ re-enters the Hubble horizon, is obtained by using the entropy conservation to be
\begin{align}\label{eq:at}
\frac{a(M_{H})}{a_0 (=1)} = \left( \frac{g_{*,s}(T_0)}{g_{*,s}(T(M_{H}))} \right)^{1/3} \frac{T_0}{T(M_{H})}\ ,
\end{align}
where $T_{0}=2.725\mathrm{K}$ denotes the present temperature of the \ac{CMB}, and the temperature $T(M_{H})$ is given by the Friedmann equation, i.e.,
\begin{align}\label{eq:ht}
3 H^2(M_{H}) M_{\text{G}}^2= \rho \approx \rho_{\text{rad}} = \frac{\pi^2 g_{*,\rho} (T(M_{H}))}{30} T^4(M_{H})\ ,
\end{align}
where $M_{\text{G}} = M_{\text{P}}/\sqrt{8\pi}$ is the reduced Planck mass. 
The relation between the horizon mass $M_{H}$ and the Hubble radius $H^{-1}$ is given by 
\begin{align}\label{eq:defmhh}
M_{H} =& \frac{ 4 \pi}{3}  \(H(M_{H})\)^{-3} \rho\ .
\end{align}
Combining Eq.~(\ref{eq:defmhh}) with the left equality of Eq.~(\ref{eq:ht}), we  have the following formula between $H$ and $M_H$, i.e.,
\begin{align}\label{eq:hm}
H(M_{H})=& 4 \pi  \frac{M_{\text{G}}^2 }{M_{H}}\ .
\end{align}
By combining Eq.~(\ref{eq:hm}) with the right equality of Eq.~(\ref{eq:ht}), we thus obtain a relation between $M_{H}$ and $T$, i.e.,
\be\label{eq:mht}
M_{H}=12\({\frac{10}{g_{\ast,\rho}(T)}}\)^{1/2}\frac{M_{G}^{3}}{T^{2}}\ .
\ee
Combining Eqs.~(\ref{eq:defkah}),~(\ref{eq:at}),~(\ref{eq:hm}) and (\ref{eq:mht}) 
 together, we obtain
\begin{align}\label{eq:kmh}
\frac{k}{k_*} = 7.49 \times 10^7  
\left( \frac{M_\odot}{M_{H}} \right)^{1/2} \left( \frac{g_{*,\rho} (T(M_{H}))}{106.75} \right)^{1/4} \left( \frac{g_{*,s} (T(M_{H}))}{106.75} \right)^{-1/3} \ ,
\end{align}
where $M_\odot$ denotes the solar mass and $k_*=0.05 \text{Mpc}^{-1}$.

\begin{figure}[tbh]
 \centering
\includegraphics[width=0.53 \columnwidth]{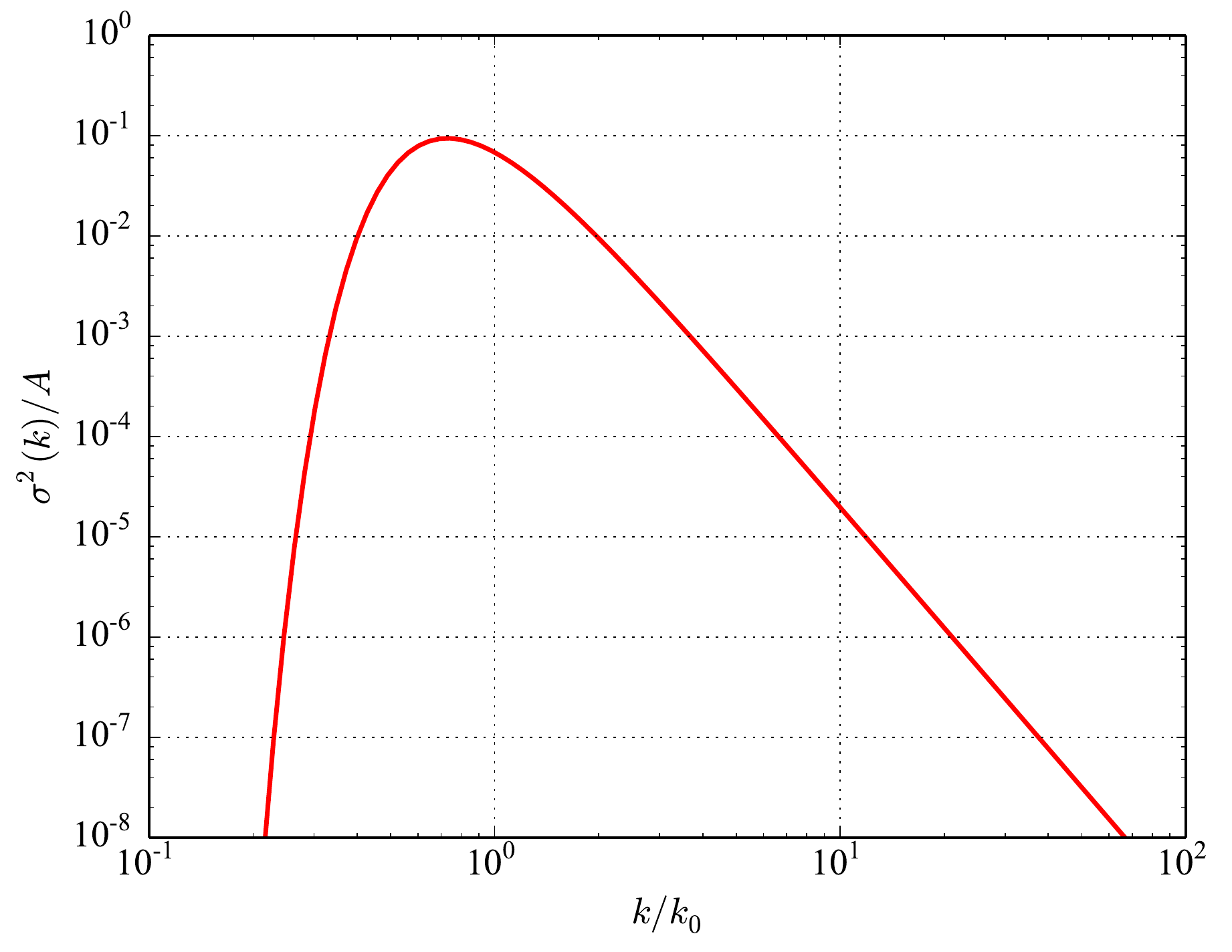}
  \caption{Coarse grained density function calculated by assuming $P_{\zeta}(k)$ to be a delta function of  $\ln k$.}
 \label{fig:sigmak}
\end{figure}

In the following, we depict Fig.~\ref{fig:sigmak} to show $\sigma^{2}(k)/A$ versus $k/k_{0}$. In the wavenumber space, the peak of the coarse grained perturbations shifts from the original peak $k_0$. Numerically, the shifted peak is obtained as $k=0.730715k_0$. 
\be \label{eq:peak_shift}
\frac{k_{\mathrm{(peak~of~PBHs)}}}{k_{\mathrm{(peak~of~primordial~curvature~perturbations)}}}=0.730715\ .
\ee
In our example by assuming $P_{\zeta}(k)$ to be the delta function,
the wavenumber in the denominator is nothing but $k_0$.  When a
\ac{PBH} forms, the shorter scales had already experienced the radiation
pressure and have been smoothened. Therefore, the \ac{PBH} mass scale
corresponds to the coarse grained perturbation scale.  In other words,
$k$ in Eqs.~(\ref{eq:k-m}) and \eqref{eq:kmh} should be the one
appearing in the numerator of the left-hand side of
Eq.~\eqref{eq:peak_shift}.
%%

%%%%%%%%%%%%%%%%%%%%%%%%%%%%%%%%%%%%%%%%%%%%%%%%%%%%%%%%%%%%%%%%%%%%%%
\section{Formalism for the merger rate of \ac{PBH} binaries}
%%%%%%%%%%%%%%%%%%%%%%%%%%%%%%%%%%%%%%%%%%%%%%%%%%%%%%%%%%%%%%%%%%%%%%
\label{append:mr}
\noindent

Given the fraction of \acp{PBH} in CDM,
namely $f_{\mathrm{PBH}}$~\footnote{As was discussed in the Introduction, we
  use a monochromatic mass distribution of \acp{PBH} as a reasonable
  approximation here. Therefore, we use the \ac{PBH} abundance
  $f_{\mathrm{PBH}}$ instead of the \ac{PBH} mass function $f(M)$.
},
for a fixed \ac{PBH} mass $M$, the probability that a \ac{PBH} binary coalesces within the cosmic time interval $(t,t+dt)$ is given by
(see e.g. Refs.~\cite{Sasaki:2016jop,Sasaki:2018dmp} for details)
\begin{align}
	dP_t=
	\begin{cases}
		\frac{3}{58}\left[-\left(\frac{t}{t_0}\right)^{\frac{3}{8}}+\left(\frac{t}{t_0}\right)^{\frac{3}{37}}\right]\frac{dt}{t}\quad\textrm{for}~ t< t_c\\
		\frac{3}{58}\left(\frac{t}{t_0}\right)^{\frac{3}{8}}\left[-1+\left(\frac{t}{t_c}\right)^{-\frac{29}{56}}\(\frac{4\pi}{3}f_{\mathrm{PBH}}\)^{-\frac{29}{8}}\right]\frac{dt}{t}\quad\textrm{for}~ t\geq t_c,
	\end{cases}
	\label{dpt}
\end{align}
where we define $t_0=({3}/{170})\{{\bar{x}}^4/[ (G M)^3 \(4\pi f_{\mathrm{PBH}}/3\)^{4}]\}$ and $t_c =t_0 \(4\pi f_{\mathrm{PBH}}/3\)^{37/3}$, and $\bar{x}=[3M/\(4\pi \rho_{\mathrm{PBH,eq}}\)]^{1/3}$ is the physical mean separation of \acp{PBH} at the epoch of matter-radiation equality. Here $\rho_{\mathrm{PBH,eq}}$ is the energy density of the \acp{PBH} at the epoch of matter-radiation equality.
Multiplying $dP_{t}/dt$ by the present average number density of \acp{PBH}, one can obtain the merger rate of \ac{PBH} binaries as 
\begin{align}
	R_{\textrm{PBH}}(z)=\(\frac{f_{\textrm{PBH}}\Omega_{\textrm{CDM}}\rho_{\mathrm{c}}}{M}\)\frac{dP_t}{dt}\ .
	\label{merger rate}
\end{align}
The redshift $z$ is related to the cosmic time $t$ through $t=\int_{z}^{\infty}dz^{\prime}/[(1+z^\prime)H(z^\prime)]$, where $H(z)=H_{0}[\Omega_{r,0}(1+z)^{4}+\Omega_{m,0}(1+z)^{3}+\Omega_{\Lambda}]^{1/2}$ is Hubble parameter at redshift $z$. The quantities $\Omega_{r,0}$ and $\Omega_{m,0}$ denote the present energy-density fractions of radiations and non-relativistic matter, respectively. The present energy-density fraction of dark energy is derived as $\Omega_{\Lambda}=1-\Omega_{r,0}-\Omega_{m,0}$.
Here $\rho_{\mathrm{c}}=3H_{0}^{2}M_{G}^{2}$ is the critical energy density of the Universe.
Throughout this paper, we adopt the $\Lambda$CDM model with cosmological parameters measured by Planck satellite \cite{Ade:2015xua}.

%%%%%%%%%%%%%%%%%%%%%%%%%%%%%%%%%%%%%%%%%%%%%%%%%%%%%%%%%%%%%%%%%%%%%%
\section{Energy spectrum of gravitational waves}
%%%%%%%%%%%%%%%%%%%%%%%%%%%%%%%%%%%%%%%%%%%%%%%%%%%%%%%%%%%%%%%%%%%%%%
\label{append:d}
\noindent
In the non-spinning limit, the inspiral-merger-ringdown energy
spectrum for a \ac{BBH} coalescence takes the following form
\cite{Ajith:2007kx,Ajith:2009bn} 
\be
\frac{dE_{\mathrm{GW}}}{d\nu_{\mathrm{s}}}\(\nu_{\mathrm{s}}\)=\frac{\(G\pi\)^{2/3}M_{c}^{5/3}}{3}
\begin{cases}
\nu_{\mathrm{s}}^{ -1/3}\quad\mathrm{for}~\nu_{\mathrm{s}}<\nu_1 \\
w_1 \nu_{\mathrm{s}}^{2/3}\quad\mathrm{for}~\nu_1\le \nu_{\mathrm{s}}<\nu_2\\
w_2\frac{\sigma^{4}\nu_{\mathrm{s}}^{2}}{\(\sigma^{2}+4\(\nu_{\mathrm{s}}-\nu_{2}\)^{2}\)^{2}}\quad\mathrm{for}~\nu_2\le \nu_{\mathrm{s}}\le \nu_3\\
0\quad\mathrm{for}~\nu_3 < \nu_{\mathrm{s}}
\end{cases}
\ee
where $\nu_{s}$ is a \ac{GW} frequency in the source frame, $w_1$ and $w_2$ are two normalization constants that make the spectrum to be continuous. 
The parameters $\nu_i$ ($i=1,2,3$) and $\sigma$ can be expressed in terms of $M_{t}$ and $\eta$ as follows
\be
&&\pi M_{t} \nu_1=(1-4.455+3.521)+0.6437\eta-0.05822\eta^2-7.092\eta^3\\
&&\pi M_{t} \nu_2 = (1-0.63)/2+0.1469\eta-0.0249\eta^2+2.325\eta^3 \\
&&\pi M_{t} \sigma = (1-0.63)/4 -0.4098\eta +1.829\eta^2-2.87\eta^3 \\
&&\pi M_{t} \nu_3 = 0.3236 -0.1331\eta -0.2714\eta^2 +4.922\eta^3
\ee
which can be found in Table~1 of Ref.~\cite{Ajith:2009bn}.  Here
$M_{c}$ is the chirp mass, i.e.,
$M_{c}^{5/3}=m_{1}m_{2}(m_{1}+m_{2})^{-1/3}$, and $M_{t}=m_{1}+m_{2}$
is the total mass. The symmetric mass ratio is defined by
$\eta=m_{1}m_{2}(m_{1}+m_{2})^{-2}$, which gives $0.25$ in this work, 
since we assume the monochromatic mass of \acp{PBH}.
The cutoff frequency is given  to be $\nu_{\mathrm{cut}}=\nu_{3}$.

%%%%%%%%%%%%%%%%%%%%%%%%%%%%%%%%%%%%%%%%%%%%%%%%%%%%%%%%%%%%%%%%%%%%%%
\section{Curvature-induced gravitational waves in a nutshell}
%%%%%%%%%%%%%%%%%%%%%%%%%%%%%%%%%%%%%%%%%%%%%%%%%%%%%%%%%%%%%%%%%%%%%%
\label{append:b}
\noindent
We briefly summarize the semi-analytic calculation of the \ac{SGWB} spectrum induced in the \ac{RD} era from the non-linear (tensor-scalar-scalar) mode coupling. The details are described in Ref.~\cite{Kohri:2018awv} and references therein.
The energy-density fraction of the induced \ac{SGWB} is given by
\be\label{eq:isgwb}
\left. \Omega_{\mathrm{GW}}(k)\right |_{\mathrm{today}}=\frac{\Omega_{\mathrm{r},0}}{24}\(\frac{g_{\ast,\rho}(T)}{g_{\ast,\rho}(T_{\mathrm{eq}})}\)\(\frac{g_{\ast,s}(T)}{g_{\ast,s}(T_{\mathrm{eq}})}\)^{-4/3}\(\frac{k}{aH}\)^{2}\overline{\mathrm{P}_{h}(\tau,k)}\ ,
\ee
where cosmic temperature $T=T(M_{H}(k))$ with the horizon mass
$M_{H}(k)$, $a H$ and conformal time $\tau$ are to be evaluated at (a
time somewhat after) the horizon entry of the relevant mode (when the
$\Omega_{\mathrm{GW}}$ has reached a temporary asymptotic value).  In
fact, $T(M_{H}(k))$ can be numerically evaluated by combining
Eq.~(\ref{eq:mht}) with Eq.~(\ref{eq:kmh}).  The last two factors in the
above formula are given by 
\be\label{eq:pst}
\(\frac{k}{aH}\)^{2}\overline{\mathrm{P}_{h}(\tau,k)}=4 \int_{0}^{\infty}d v \int_{-|1-v|}^{1+v}d u \[\frac{4v^2 - (1+v^2-u^2)^2}{4 u v}\]^{2}(k\tau)^{2}\overline{I^{2}(v,u,k\tau\gg 1)}\mathcal{P}_{\zeta}(kv)\mathcal{P}_{\zeta}(ku)\ .
\ee
In the above equation, we have  
\be
\label{eq:i2}
(k\tau)^{2}\overline{I^{2}(v,u,k\tau\gg1)}&=& \frac{1}{2} \left( \frac{3(u^2+v^2-3)}{4 u^3 v^3 } \right)^2 \left( \left( -4uv+(u^2+v^2-3) \ln \left| \frac{3-(u+v)^2}{3-(u-v)^2} \right| \right)^2  \right. \nonumber \\
&&  \left. \qquad \qquad \qquad \qquad \qquad   + \pi^2 (u^2+v^2-3)^2 \Theta ( v+u-\sqrt{3}) \right).
\ee
Then, we combine the above three equations to compute the
energy-density spectrum of the induced \ac{SGWB} in this work.  In
case of the delta function source (Eq.~\eqref{eq:pzeta}), the integral
is easily calculated to obtain Eq.~\eqref{eq:isgwbppp}.

\bibliography{sgwb_pbh-3}
%%%%%%%%%%%%%%%%%%%%%%%%%%%%%%%%%%%%%%%%
%%%%%%%%%%%%%%%%%%%%%%%%%%%%%%%%%%%%%%%%
\end{document}